\documentclass[letterpaper, 10 pt, conference]{IEEEtran}
\ifCLASSOPTIONcompsoc
  \usepackage[nocompress]{cite}
\else
  \usepackage{cite}
\fi

\ifCLASSINFOpdf
  \usepackage[pdftex]{graphicx}
  \DeclareGraphicsExtensions{.pdf,.jpeg,.png}
\else
  \usepackage[dvips]{graphicx}
  \DeclareGraphicsExtensions{.pdf,.jpeg,.png}
\fi

\usepackage{hhline}

\usepackage{ifthen}
\usepackage{color}
\usepackage{amssymb}

\newboolean{showcomments}
\setboolean{showcomments}{false} % toggle to show or hide comments
\ifthenelse{\boolean{showcomments}}
  {
		\newcommand{\nbb}[2]{
		\fcolorbox{black}{yellow}{\bfseries\sffamily\scriptsize#1}
		{\sf$\blacktriangleright$\textcolor{blue}{\textit{#2}}$\blacktriangleleft$}
		}
		
		\newcommand{\remarks}[1]{\color{red}[#1]\color{black}}

		\newcommand{\del}[1]{\textcolor{red}{\sout{#1}}} % please delete
  }
  {
		\newcommand{\nbb}[2]{}
		\newcommand{\remarks}[1]{}

		\newcommand{\del}[1]{} % please delete
  }

\usepackage{tabularx}

\usepackage{url}
\usepackage{enumitem}
\usepackage{textcomp}

\usepackage{tikz}

\newcommand\copyrighttext{%
  \footnotesize \textcircled{c} 2017 IEEE International Conference on Software Architecture (ICSA 2017), Gothenburg, Sweden \\%
Giaimo, Federico, and Berger, Christian. "Design Criteria to Architect Continuous Experimentation for Self-Driving Vehicles." \textit{Software Architecture (ICSA), 2017 IEEE International Conference on}, IEEE, 2017, DOI: 10.1109/ICSA.2017.36, \url{http://ieeexplore.ieee.org/abstract/document/7930218/}}
\newcommand\copyrightnotice{%
\begin{tikzpicture}[remember picture,overlay]
\node[anchor=south,yshift=10pt] at (current page.south) {\fbox{\parbox{\dimexpr\textwidth-\fboxsep-\fboxrule\relax}{\copyrighttext}}};
\end{tikzpicture}%
}

\begin{document}
\title{Design Criteria to Architect Continuous\\Experimentation for Self-Driving Vehicles} 

\author{\IEEEauthorblockN{Federico Giaimo}
\IEEEauthorblockA{Department of Computer Science and Engineering\\
Chalmers University of Technology\\
giaimo@chalmers.se}
\and
\IEEEauthorblockN{Christian Berger}
\IEEEauthorblockA{Department of Computer Science and Engineering\\
University of Gothenburg\\
christian.berger@gu.se}}

\IEEEtitleabstractindextext{%
\begin{abstract}
The software powering today's vehicles surpasses mechatronics as the dominating engineering challenge due to its fast evolving and innovative nature.
%Source: http://www.kurzweilai.net/googles-self-driving-car-gathers-nearly-1-gbsec?utm_source=datafloq&utm_medium=ref&utm_campaign=datafloq
In addition, the software and system architecture for upcoming vehicles with automated driving functionality is already processing \texttildelow 750MB/s -- corresponding to over 180 simultaneous 4K-video streams from popular video-on-demand services.
Hence, self-driving cars will run so much software to resemble ``small data centers on wheels'' rather than just transportation vehicles.

%CBe: Source: https://blogs.bing.com/search-quality-insights/2013/08/08/large-scale-experimentation-at-bing
Continuous Integration, Deployment, and Experimentation have been successfully adopted for \textit{software-only} products as enabling methodology for feedback-based software development. 
For example, a popular search engine conducts \texttildelow 250 experiments each day to improve the software based on its users' behavior.

This work investigates design criteria for the software architecture and the corresponding software development and deployment process for complex cyber-physical systems, with the goal of enabling Continuous Experimentation as a way to achieve continuous software evolution. 
Our research involved reviewing related literature on the topic to extract relevant design requirements.

The study is concluded by describing the software development and deployment process and software architecture adopted by our self-driving vehicle laboratory, both based on the extracted criteria.
\end{abstract}

% keywords chosen from the official taxonomy (http://www.ieee.org/documents/taxonomy_v101.pdf)
}

\maketitle

\IEEEdisplaynontitleabstractindextext

\copyrightnotice

\section{Introduction}
\label{sec:introduction}
Due to the ever-increasing omnipresence of cyber-physical systems and the corresponding growth in complexity of its software, reliable and well-performing software architecture and corresponding processes are essential to their success. 
This holds true when considering complex and interconnected cyber-physical systems like self-driving vehicles, due to the compelling safety implications for both the users and the people in their surroundings. 
The architecture plays a fundamental role when taking into account software evolution and maintenance as well, since they require well-established and reliable practices to guide software change to minimize development efforts without jeopardizing safety aspects and to preserve traceability. 

In the automotive context, if we consider the general situation, the possibility of an emerging software fault would force the manufacturer to issue very costly vehicle recalls and possibly face investigations (as it happened for example for the Toyota's unintended acceleration case~\cite{URL_toyota}).
This is due to the fact that the software controlling commercial vehicles is for the most part related to safety-critical aspects, and with few exceptions these vehicles generally lack the possibility to receive Over-The-Air (OTA) critical software updates, which would allow to solve the problem directly in the deployed systems. 

\begin{figure}[t]
\begin{center}
\hspace*{-0.4cm}\includegraphics[width=9.3cm]{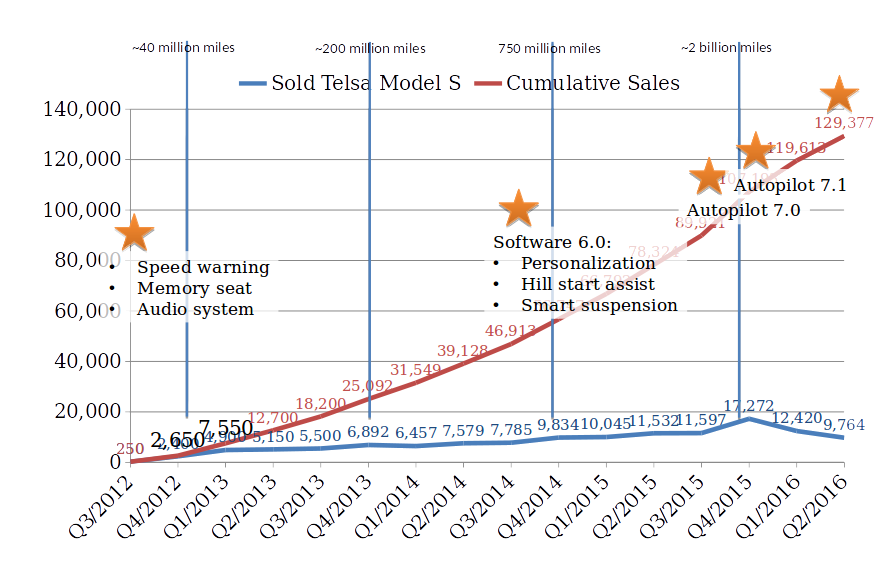}
\caption{Per-quarter and cumulative sales of Tesla Motor's Model S, and corresponding cumulative distance driven (vertical lines) from mid-2012 to mid-2016. At the end of 2015 the company sold more than 100 thousands cars and accumulated around 2 billion driven miles. The stars represent software releases.
}
\label{fig:tesla}
\end{center}
\end{figure}

The strategy of some recent car manufacturers, for example Tesla~Inc., is an exception to this picture, as it seems to have taken these considerations into account. 
As they reported, a step-by-step approach was applied to roll-out their ``Autopilot'' software: the company developed the car's Advanced Drivers Assistance Systems (ADAS) and gradually and incrementally turned it into a ``supervised'' self-driving feature via OTA software updates~\cite{URL_tesla} that were developed by the company.
The massive amount of data collected by the cars themselves along the years (as depicted in Fig.~\ref{fig:tesla}) was presumably used as additional test and validation data, to complement the already significant testing procedures that are in place for automotive software, e.g.~software test suites, Hardware-In-Loop simulation, ``Test farms''~\cite{VW16}.

In less demanding software contexts like e-commerce environments, it is possible to tackle the need for software improvement by adopting Continuous practices. 
These are a set of well-known Extreme Programming (XP) practices proposed to improve the development process; they include Continuous Integration, Delivery, Deployment, and Experimentation. 
Continuous Integration (CI) is the practice of integrating the developers' work in the code base several times a day in order to decrease the probability of facing integration conflicts.
Continuous Delivery is the practice of producing software in smaller cycles so that it could be released at any time. 
Continuous Deployment (CD) goes one step further, meaning that whenever the software is ready to be released and delivered it should be done, thus deploying the software to the units every time it is possible.
Lastly, Continuous Experimentation (CE) is a recognized and increasingly adopted practice, especially in the context of software-intensive web-based platforms, that enables product developers to perform controlled post-deployment experiments to collect significant data and statistics, as outlined by Fagerholm et al.~\cite{FGMM16}.
These experiments can be for example new minimum viable features (software features containing or performing just the bare core of its purpose, deployed to gain feedback for later development and expansion) or software patches that are deployed and activated in a subset of all the available systems, not necessarily with the users being aware of it.
This practice's final goal is to use the collected data to assess issues and to drive the development of new features, thus allowing the development process to base more informed decisions on the factual evidence collected ``in the field''.
The enabling factors for this process are the data collection capabilities and both the \textit{downlink} and the \textit{uplink} connections to the deployed systems, as to perform CE means to deploy additional or replacement software to be run and then to receive back data from the deployed units.

\subsection{Problem Domain \& Motivation}

In the case of safety-critical cyber-physical systems there are still many challenges to be overcome in order for Continuous Experimentation to be adopted, as shown in our previous study~\cite{GYBC16}.
These challenges are mostly related to the constraints in terms of resources and safety-critical regulations that the software must always obey. 
They can be summarized with: 
\begin{itemize}
\item hardware constraints, industries usually provide automotive platforms with ``just enough'' processing power due to the costs of mass production, which means that it can be hard to find the additional computation/memory/bandwidth that is needed for additional software to run in ``piggyback'' mode on a car if it was not built for that task;
\item 
safety constraints, the software must guarantee that the performances are always compliant with safety regulations, otherwise the car is considered illegal on public roads; 
\item scalability of data transfer, as the sensors used for self-driving vehicles generate a vast amount of data each second (a test run with our vehicles can produce around 300MB/s, mostly due to camera and radars' data streams), a sound strategy is needed to manage how the vehicles will upload their data back to the manufacturer's research facilities when needed, especially considering that these vehicles may be present in great numbers in metropolitan areas. 
Even now that autonomous vehicles are still transitioning from being just prototypes to being a reality, the numbers are going from the 100 Volvo's \textit{Drive Me} trial vehicles to be released in 2017~\cite{URL_volvo} to the thousands of sold Tesla Model S (Fig.~\ref{fig:tesla}).
\end{itemize}

Another significant reason to strive for an architecture that can enable Continuous Experimentation is to support vehicle research not just in the sense of evolving software for one platform, but also considering the context of several heterogeneous vehicles powered by the same software family. 
Our university's vehicle laboratory is called Revere~\cite{URL_revere}, and it is currently hosting a SUV (Volvo XC90), a truck (Volvo FH16) which participated to the Grand Cooperative Driving Challenge that took place in May 2016 in the Netherlands~\cite{URL_gcdc2011}, an active-steered truck dolly, and a number of miniature vehicles for educational purposes.
In order to minimize future maintenance or refactoring efforts for the software that needs to operate on such different platforms, the software process and architecture needs to be able to scale and facilitate future change and evolution. 
It is not difficult to transpose this need from the context of a modern but small-sized vehicle laboratory to the much larger scale of an industrial manufacturer who envisions several self-driving vehicle models. 

\subsection{Research Goal}
The goal of this work is to find properties of the software architecture and process required to enable Continuous Experimentation for a complex cyber-physical system such as an autonomous vehicle.
The vehicle runs a distributed software, which has to provide the necessary guarantees for the execution of real-time safety-critical functions.
We identified the following research questions:
\begin{enumerate}[leftmargin=*,labelindent=1em,label=$RQ{\arabic*}:$]
\item What are functional and non-functional requirements for a software development and deployment process to support Continuous Experimentation targeting self-driving vehicles? 
\item What are requirements for the software architecture to support Continuous Experimentation on self-driving vehicles?
\end{enumerate}

These questions will be addressed by means of reviewing relevant literature in a structured way in order to extract the requirements of interest.

\subsection{Structure of the Document}
The document is structured as follows: in Section~\ref{sec:related} other selected works in literature related to software architecture for autonomous vehicles are summarized; in Section~\ref{sec:requirements} the extracted functional and non-functional requirements for the software process and software architecture are described; in Section~\ref{sec:software} the software process and architecture adopted in our self-driving vehicles' laboratory Revere are presented; in Section~\ref{sec:discussion} the results are discussed and the threats to validity are reported; lastly, in Section~\ref{sec:conclusions}, the present work is concluded and directions for future works are outlined.

\section{Related Work}
\label{sec:related}
Relevant works in literature covering architectures for self-driving vehicles have been searched for in the digital libraries provided by IEEE and ACM using the key phrases ``\textit{architecture} AND \textit{`autonomous vehicle'}~'' and ``\textit{architecture} AND \textit{`autonomous driving'}~'', and considering only the works published after 2006 to include the results from and inspired by the 2007 DARPA Urban Challenge. 
The combined results from the IEEE Xplore Digital Library and from the ACM Digital Library amounted to 223 articles and can be found at the link: \url{https://se-div-c3s-1.ce.chalmers.se:7001/fbsharing/vNPuDpfP}. % we got 223 papers (IEEE=194,ACM=29)
As most of the literature recognizes the advantage of a distributed architecture, the majority of the works are not hereby summarized or mentioned due to the fact that they do not introduce new elements to the discussion, although they are still relevant. 

One first relevant work is Baker and Dolan~\cite{BD08}, that describes the software powering in the autonomous car ``Boss'', the vehicle that won the 2007 DARPA Urban Challenge.
In this paper it is outlined how the software was built following the \textit{Observer} pattern, which works in a conceptually similar way as the more common \textit{Publish/Subscribe} pattern, providing both module inter-communication and decoupling. 

One of the recent works on architectural aspects is Behere and T{\"o}rngren~\cite{BT16}, which focuses on the functional perspective of the architecture itself.
This paper highlights the differences between the case where the high-level logic and the low-level platform are integrated and the case where the autonomy logic and the platform were treated as separated but collaborating units.
According to the authors, as much as it could be tempting to achieve the former situation with the two macro-levels (logic and platform) closely interacting to achieve autonomy, that is almost always impracticable for those who are not an Original Equipment Manufacturer (OEM), due to the vast knowledge of the hardware platform that becomes necessary. 
Furthermore they state that the coupling between intelligence and platform needs to be carefully managed in order to avoid side-effects or unclear separation of scopes in the software, while on the contrary the architectures in which the two macro-levels are treated as separated are the most obvious for research organizations that work on augmenting an existing vehicle to provide it with autonomous functionality. 
One of this article's conclusions is that these logically separated architectures are the ones that give the most guarantees regarding simplification of modeling and development process, achieve the best separation of concerns among the constituting software modules, and allow for better software testing and reuse. 

Another related work focusing on the architectural level was discussed in our previous work~\cite{BD14}, where the results of a systematic literature review and a multiple case study are presented. 
The authors summarized their findings highlighting the following key aspects that characterized the resource-constrained system and software architectures for self-driving vehicles they analyzed:

\begin{itemize}
\item a strict separation between low-level and high-level functions, as it allows to more easily develop solutions for very diverse tasks, e.g. hardware control, situation awareness, etc.;
\item hardware abstraction layer over the physical devices, in order to decouple the high-level software development from the concrete hardware, and also to allow to undertake hardware replacements/upgrades without the need to rewrite software;
\item loosely coupled components, due to the software being distributed and its modules communicating with one another via TCP/UDP message-passing techniques or a service registry;
\item platform-independent data representation, as the software modules used aggregated custom messages to communicate;
\item platform-independent development process, using for example cross-compiling tool chains.
\end{itemize}

%%%%%%%%%%%%%%%%%%%%%%%%%%%%%%%%%%%%%%%%%%%%%%%%%%%%%%%%%%%%%%%%%%%%%%%%%%%%%%%%%%%%%%%%%%%%%%%%%%%%%%%%

\section{Functional and Non-Functional Properties}
\label{sec:requirements}
The functional and non-functional properties extracted from the listed articles will now be described. The properties can have general validity and purposes, but the main focus of this study is to identify those that are closest to Continuous practices, especially Continuous Delivery/Deployment and Continuous Experimentation.

The summarized papers often relate to urban traffic contexts, from simulated urban scenarios as it is in the case of autonomous cars competitions, to actual urban environments.
Some recurrent functional properties can be extracted for those vehicles that are re-engineered to be autonomous, such as access to perception sensors and systems (FR1) in order to perform localization, lane detection, and obstacle detection tasks; and access to full vehicle control (FR2), including steering, speed management, and gear shifting. 

One commonly shared functionality, which is necessary for obtaining results out of the deployed experiments, is the ability to log internal activity and other relevant metrics by instrumenting the software (FR3) for later inspection.
Additional functional requirements can vary widely depending on the final application setting, as for example some vehicles may be specifically designed for moving in rural areas, some may be focusing on competitions where only selected scenarios can occur, and so on. 
For this reason, the task to find common functional properties for all architectures is generally unlikely to succeed as it is very much related to the final goal of each project.

Considering instead additional criteria to fulfill in order to enable Continuous Experimentation, the key requirements so far missing are the enabling of data transmission from the developers to the deployed system (FR4) and the feedback loop in the opposite direction (FR5).
These can be achieved by providing the vehicle with a network interface and including in the software a module or component acting as a ``communication gateway'' between the internal communication network and the external server, with the goal to filter those internal messages that need to be relayed back to the developers, and to forward in the internal bus those messages from the external server that need to be processed by the system. 
Remote data exchange is however a very delicate topic as it may enable the possibility for third parties to secretly hack into the system for undesired monitoring or other more harmful purposes. 
One proposed way to contrast this threat is to keep vehicles offline as much as possible~\cite{URL_avoidinternet}, and enable secured communications only when deemed necessary. 
% 5 FR in total

Non-functional requirements are instead more general and more related to the adopted architectures, and from our literature study some recurrent ones will be highlighted, such as reliability (NFR1), testability (NFR2), safety (NFR3), scalability (NFR4), and separation of concerns (NFR5), which implies abstraction layers between hardware and software and between data and exchanged messages. 
Reliability is usually envisioned through the adoption of health checking techniques and modular software architectures which are able to limit faults propagation.
Moreover, the internal communication layer in the vehicles are additionally subjected to reliability constraints in terms of packet loss and latency.
The reliability property is very much connected with the testability of its software and technologies, which is usually achieved by enabling the software to be run in complex simulations, Software/Hardware-In-Loop test benches, and ``Test farms''.
Safety is of course a fundamental requirement to be fulfilled for all vehicles. 
Rigid guidelines and standards for automotive software are in place to assure that it is as much free from side-effects as possible, and to ensure reliability and testability, which in turn guarantee that the safety features deployed to the vehicles would not fail when needed.

Despite the limitation posed on some advanced programming language features, e.g.~the unavailability of dynamic memory allocation, the automotive software's complexity and size are ever-increasing, as it is described by Broy~\cite{Broy06}, meaning that more and more sophisticated functions are available to monitor both the vehicles' internal status and its surroundings, in the effort to prevent the driver to find herself in a hazardous situation.
For autonomous vehicles, one of these features is the ``emergency button'': the mechanism in place available to the passengers to interrupt the autonomic algorithms, thus regaining immediate manual control of the platform.
Scalability comes often into play for autonomous vehicles due to the vast amount of data that they have to collect and process in order to make sense of the world around them.
An architecture that is not scalable would risk the impossibility to add new sensors or hardware platforms to their internal network, or the chance of not being able to process all the needed data in particularly stressful scenarios.
Keeping concerns separated instead means that the software modules are connected only through well-defined message interfaces, and are not correlated in other ways with each other.
This helps to avoid inter-module relations and increases the overall software's modularity and maintainability.

Additional desirable non-functional requirements connected with the software development process are the simplicity to involve new developers (NFR6) as more than just one software team are usually involved in the software production phase; facilitation for operators (NFR7), meaning that the software should not be hard to operate for those who are not developers, thus including testers and end-users. 

Lastly, a short cycle from development to deployment (NFR8) is necessary whenever possible in order to roll out changes and new features in a timely manner, for this reason it is instrumental for both Continuous Deployment and Experimentation.
However it should be considered that inherently mobile platforms such as vehicles could experience prolonged periods of time incapable of connecting to their software server, thus impairing at least temporarily the possibility to timely receive software updates.

It is worth noting that requirements like reliability, scalability, separation of concerns, facilitation for operators, short deployment cycle, logging and data communication between the systems and the developers are also shared by data centers.
This consideration is in line with the way software in cars has increased in size and complexity, and how it is likely to further increase in the future to accommodate for the necessarily more advanced features and algorithms that will enable fully autonomous driving capabilities.
Considering the future challenges related to the complexity of needed algorithms, the amount of data that will need to be processed in autonomous vehicles and their real-time requirements to guarantee safety of operations, it is already possible to foresee the emergence of similarities between the two fields.

% 8 NFR in total
%%%%%%%%%%%%%%%%%%%%%%%%%%%%%%%%%%%%%%%%%%%%%%%%%%%%%%%%%%%%%%%%%%%%%%%%%%%%%%%%%%%%%%%%%%%%%%%%%%%%%%%%

\section{Revere's Software Architecture and Development \& Deployment Process}
\label{sec:software}

The discussion about the software adopted in the Revere laboratory will be separated in \textit{process} and \textit{architecture}.
The \textit{software process} relates to the way the software development and deployment process is structured, and how the software itself is then deployed to the system. 
The \textit{software architecture} section will instead describe the way the software modules are themselves interacting with each other, and how communication is structured.
Both process and architecture were designed with the aforementioned requirements in mind.

\subsection{Process}

In order to streamline the software development process towards the production phase and to fulfill non-functional requirements related to developers and operators, a container-based approach has been adopted, meaning that the build environment is encapsulated in a structured stack of \textit{Docker}~\cite{URL_docker} ``images'' (layers of the stack) separated according to the provided functionality, containing sources and dependencies of the software to be built based on the adopted middleware; this implies that the software is always built in a deterministic, reproducible and traceable environment. 
Each layer thus encapsulates various functions and aspects for a self-driving vehicle in order to separate concerns and to introduce abstraction layers (NFR5). 

\begin{figure}[t]
\begin{center}
\hspace*{-0.2cm}\includegraphics[width=9.2cm]{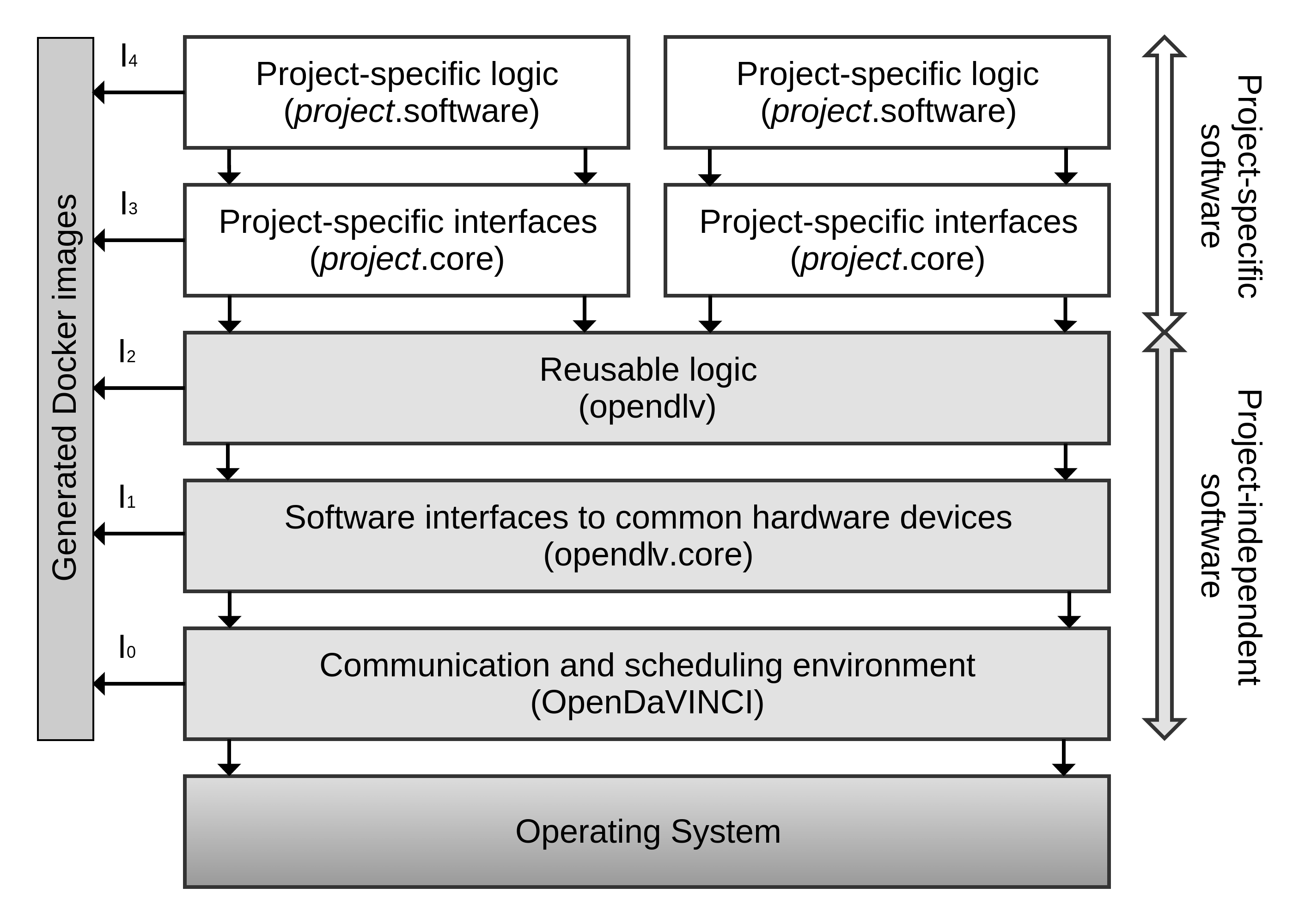}
\caption{Software stack adopted by the Revere self-driving vehicle laboratory. Each layer generates a docker image providing a well-specified set of logical functions.} 
\label{fig:layered}
\end{center}
\end{figure}

Acting as foundation at the lowest layer of the entire software stack, the scheduling and communication environment OpenDaVINCI~\cite{URL_opendavinci} enables a uniform distributed software environment for all running modules and allows real-time, deterministic and traceable scheduling of tasks. 
On top of this basic layer all the open source software interfaces to the so far supported hardware components are collected in another software layer called ``opendlv.core'', they include for example access to cameras, laser rangefinders, and GPS units.
To ease reuse among different projects, on top of this layer another one called ``opendlv'' is placed: it contains reusable logic, handling for example coordinate transformation or data visualization. 
As it is shown in Fig.~\ref{fig:layered}, the software stack up until the opendlv layer acts as a common ground for the project-individual stacks on top of it, as the higher layers become more project-specific to cope with less general needs originating, for example, from specific hardware settings or non-disclosure agreements to protect intellectual property.
As an example, further software layers include support to a real-size commercial truck and miniature vehicles, among the others.

The adoption of Docker images to execute the needed binaries has been proven to have a negligible impact on the system's performances, and thus the choice of whether to adopt a containerized execution environment or not can be independent of performance considerations; instead, the type of kernel that is used (Real-Time or not) was proved in our previous study~\cite{MTA+16} to be a crucial factor for the resulting performances and for this reason it should be carefully chosen.
In our setup the Volvo XC90 is operated by a computer running the ArchLinux RT kernel 4.8.15\_rt10-1.

The other significant advantage of this approach is that the only software that needs to be present on a computer to be able to develop and compile software for the project is the Docker software suite. 
This is a strategic point for our project's setting as there is more than one software team involved in development, and for this reason a solution that quickly enables a developer to work without undergoing a ``dependencies installation'' process to get started is preferred (NFR6).
When a change in the code base is needed, only the affected layer is recompiled instead of the whole software stack, thus decreasing the time needed to obtain a deployable layer (NFR8).
Not related to a specific requirement but supporting the efforts towards Continuous Integration, it is worth mentioning that after any new change the code is integrated and tested using a Jenkins server (\url{https://jenkins.io/}) before it may be considered for deployment.

After the compilation phase, the generated binaries will then be used to create a resulting layer image which is the one that will be deployed to the actual vehicles.
This image does not need to comprise the entire building environment, as the only needed items are the executable files and their dependencies.
When the resulting image is deployed to the vehicles' on-board computing platforms, the tool \textit{docker-compose} is used to start predefined applications scenarios, like e.g. \textit{autonomous drive mode} or \textit{autonomous parking mode}, based on containers spawned from the previously mentioned image. 

The tool works by simply reading a dedicated file containing all the details of the binaries that need to be run in a docker image, and then executes them in the containerized environment. 
This allows to obtain strict control over what binaries are actually started in each scenario (the number of run binaries can get to \texttildelow30 in some prototypical tests, but in real-world scenarios the total is expected to rise up to between 50 and 200, possibly on several machines), as well as to make clear their versions and configuration parameters provided when the binaries are run.
Furthermore the tools facilitates the platform usage for the operators that may or may not be knowledgeable of the platform's internal file structure and binaries location, since it allows to start the aforementioned scenarios with a single command (NFR7).

\subsection{Architecture}

Based on the related works extracted from literature, the state-of-the-art with respect to the software architecture for autonomous vehicles consists of a distributed system of software modules that guarantees scalability (NFR4), where the internal communication is founded on a message-passing protocol of choice to enable a suite of interacting services instead of a rigid and non-flexible monolithic software.
The exchanged messages are all recorded for later inspection (FR3).

Several abstraction layers are in place in order to enforce a strong separation of concerns among the actors that influence the status of the system at different levels (NFR5); the separation enables a more fine-grained control of the modules' behaviors and easier development and testing (NFR2).
Similarly, the messages that are exchanged among those same actors are well abstracted from the raw data in order to carry more relevant information and to allow for a more functional communication. 
This is achieved by adopting the open-source language ``odvd'' to define communication messages in order to overcome a typical challenge in the automotive industry with project- or even vehicle-specific data messages definition. 
This custom language was designed around the Protobuf~\cite{URL_protobuf} syntax to represent in a platform- and language-independent way the data to be transmitted, but it is independent also from the actual serialization method or version, allowing to define and exchange data messages in a transparent way for all supporting projects. 
Additionally, since its implementation is integrated in the middleware, there is no dependency from third-party libraries and future optimization is possible whenever the possibility arises. 

In our case, characteristics like reliability (NFR1) and safety (NFR3) will depend more on the high-level logic implemented in the vehicles than on the architecture \textit{per se}. 
The same can be said about the access to sensorial devices (FR1) and the vehicles' control bus (FR2).
Lastly, the software is capable of receiving and transmitting data to a predefined IP address (FR4 \& FR5).

The software, comprising the architecture and the aforementioned build tools and configuration, is freely available online as open-source projects at the following links: \url{https://github.com/se-research/OpenDaVINCI}, \url{https://github.com/chalmers-revere/opendlv.core}, \url{https://github.com/chalmers-revere/opendlv}.

\section{Discussion and Threats to Validity}
\label{sec:discussion}

\subsection{Discussion}

The results of our analysis have been summarized in Table~\ref{tab:requirements}, which contains the identified requirements, whether they relate to the software architecture or to the development process, if they are needed to support Continuous practices, which papers (belonging to the set of papers summarized in Sec.~\ref{sec:related}) referred to each requirement, and what solutions were implemented in the Revere laboratory to fulfill them. 
The aim of the analysis is to illustrate the reasoning that guided the development of our software stack, and at the same time to guide future works in the area of Continuous practices in the context of autonomous vehicles that would focus on the architectural aspects needed to achieve them. 
According to our analysis, in order to enable Continuous experimentation in a setting similar to ours, the requirements NFR7, NFR8, FR3, FR4, and FR5 will need to be fulfilled. 
This would result in deploying to the system (FR4) in short cycles (NFR8) an easy-to-use software capable of running an ``experiment'' (NFR7), of collecting the resulting data (FR3) and finally capable of reporting these data back to the developers (FR5), which in turn will be able to better choose how to steer the development efforts or which new experiment to deploy and run, thus restarting the experimentation cycle.

In the event that an experiment would introduce any risk to the status of the platform or to the safety of its passengers or other people, architectural solutions on how to counter these risks shall be envisioned.
One solution present in our platform is the emergency button that would reset the vehicle to its non-autonomous state. 
This solution is however intended as ``last resort'' and in any case relies on the presence of a human driver acting as back-up instance. 
A foreseeable more advanced approach, compatible with the vision of a truly driver-less vehicle, would for example be based on an internal ``self-preservation'' mechanism capable of stopping the deployed experimentation capabilities when some pre-defined safety aspects are violated; however, in our software this mechanism is at the present moment only theoretical, and no practical efforts have been devoted to its creation.

The software stack adopted in our laboratory and described in Sec.~\ref{sec:software} already complies to almost all of the identified requirements, being FR5 the only one that is still in its prototypical implementation. 
Additional missing pieces of the described architecture to complete the vehicle's metamorphosis from a transportation tool to a distributed ``data center on wheels'' would relate to the deployment and workload distribution of the needed binaries upon the available hardware resources, assuming the existence of a ``supervisor'' with free access to the collective pool of computational resources.
A possible step in the direction of ``distributed deploying'' could be the tool \textit{docker swarm}, which promises to provide clustering capabilities to a set of docker engines in order to virtualize them into a single virtual docker engine. 
Load balancing is an even harder topic to face, as it would imply the possibility to ``relocate'' a running module from a computational node to another in order to balance the platform's overall workload.
The challenge to be faced in order to achieve this is the capability of moving a process and its context at system's run-time to a different hardware platform, which could result in a potential disruption of critical services, even if only temporary.

The software, built with the described requirements in mind, has proved its validity in several occasions during Researcher's Days events and even as a Technical Demo at the 2016 IEEE Intelligent Vehicles Symposium, where the two Revere vehicles demonstrated a leader-following behavior.

\subsection{Threats to Validity}

Relevant threats to the validity of our results and applicability in industry are in the fundamental nature of our work: as the project is research-focused and not targeting commercial applications, some of the very strict constraints that automotive software has to abide to did not apply to our case.
Software guidelines like the ones provided by MISRA (Motor Industry Software Reliability Association)~\cite{URL_misra} and safety guidelines described in the international standard ISO 26262~\cite{URL_iso26262} were thus set aside, resulting in our software being well suited for our research applications but not ready for immediate use in a real-life context -- even though reasonable care and state-of-the-art practice has been applied during implementation like Unit testing or static code analysis.
Because of that, our test-drive phases were conducted in the privately owned proving ground AstaZero~\cite{URL_astazero}, instead of public roads.

Another threat to the validity of our work lays in the literature study, as the obtained articles were relatively few in numbers and thus relevant works may have been missed.
However, after the examination of the set of initial results, we suspect that a reason for the small number of entries is that it is not common in the context of autonomous vehicles to describe in detail the architecture's characteristics, as the most common approach is to just describe which high-level modules, e.g.~``Sensor Fusion'' or ``Trajectory Planning'', are run in the software. 

\section{Conclusions and Future Work}
\label{sec:conclusions}

This paper described relevant requirements for a software development and deployment process and software architecture that could support long-term software evolution in the context of a complex cyber-physical system such as a self-driving vehicle.
The tool of choice for the software evolution support in our study is Continuous Experimentation, one of the Continuous practices identified in Extreme Programming.
The extracted requirements have been summarized in Table~\ref{tab:requirements}.

Although being valid instruments, the adoption of Continuous practices in the cyber-physical systems field is still a relative novelty regardless of the promises that said practices carry, due to the many challenges and constraints that still limit their adoption in commercial and real-life scenarios, especially when considering safety-critical aspects.

Being established with the identified requirements in mind, the software process and software architecture adopted in our university's self-driving vehicle laboratory are finally described as a relevant proof-of-concept.

Future efforts are set to focus on the demonstration of a prototypical Continuous Experimentation iteration using the Revere laboratory's vehicles, in order to show how the entire CE process can be successfully performed on a self-driving vehicle.
Further considerations related to the type of data needed for conducting experiments as well as considerations about what functionality could -or shouldn't- be experimented on in different settings shall be likewise addressed.

\ifCLASSOPTIONcompsoc
  \section*{Acknowledgments}
\else
  \section*{Acknowledgment}
\fi
This work has been supported by the COPPLAR Project -- CampusShuttle cooperative perception and planning platform~\cite{URL_copplar}, funded by Vinnova FFI, Diarienr: 2015-04849.

\begin{table*}[t]
\renewcommand{\arraystretch}{1.5} % increase table row spacing
\caption{Summary of identified requirements, their context, if they are instrumental in supporting Continuous practices, papers that referred to them (among the one that were summarized in Sec.~\ref{sec:related}), and what solutions were implemented in our laboratory to satisfy them.}
\label{tab:requirements}
\centering
\newcolumntype{Y}{>{\centering\arraybackslash}X} % for centered column
\newcolumntype{S}{>{\hsize=.5\hsize}Y} % for smaller column
\begin{tabularx}{\textwidth}{|Y|Y|S|S|Y|}
\hline
Requirement & Context of the requirement & Continuous practices supported by the requirement & Papers referring to the requirement & Solution adopted by our Revere laboratory \\
\hhline{|=|=|=|=|=|}
NFR1: Reliability & Architecture & - & all & Managed in the software logic \\
\hline
NFR2: Testability & Architecture & - & all & The software undergoes testing phases and the logic is run in a simulated environment before being rolled out to the target system \\
\hline
NFR3: Safety & Architecture & - & all & Managed in the software logic \\
\hline
NFR4: Scalability (in terms of adopting a distributed system at all logical levels) & Architecture & - & all & The software is distributed among different computational platforms, connected through fast communication lines \\
\hline
NFR5: Separation of Concerns & Architecture & - & all & Achieved via abstraction layers among software modules and communication messages \\
\hline
NFR6: Simplicity to Involve New Developers & Process & - & none & Obtained by moving the software development phase in a containerized environment \\
\hline
NFR7: Facilitation for Operators & Process & CD/CE & none & The use cases are set up and run with a single command as they are defined in specific files \\
\hline
NFR8: Short Cycle to deployment & Process & CI/CD/CE & none & Both the software and the containerized images are built incrementally whenever possible \\
\hline
FR1: Access to Perception Devices & Architecture & - & all & Functionality provided by the OpenDaVINCI middleware \\
\hline
FR2: Access to Vehicle Control & Architecture & - & all & Functionality provided by the OpenDaVINCI middleware \\
\hline
FR3: Logging and Instrumentation & Architecture & CE & none & The OpenDaVINCI middleware can provide logging capabilities for data of interest \\
\hline
FR4: Data Transmission from Remote Server to System & Architecture & CE & \cite{BT16} & Whenever connected, the middleware can be contacted by a remote software server to receive software updates \\
\hline
FR5: Data Feedback from System to Remote Server & Architecture & CE & \cite{BT16} & Whenever connected, the middleware can upload relevant data and statistics to a remote software server \\
\hline
\end{tabularx}
\end{table*}

\ifCLASSOPTIONcaptionsoff
  \newpage
\fi

\bibliographystyle{IEEEtran}
\bibliography{ICSA17_arxiv_library}

\begin{thebibliography}{10}
\providecommand{\url}[1]{#1}
\csname url@samestyle\endcsname
\providecommand{\newblock}{\relax}
\providecommand{\bibinfo}[2]{#2}
\providecommand{\BIBentrySTDinterwordspacing}{\spaceskip=0pt\relax}
\providecommand{\BIBentryALTinterwordstretchfactor}{4}
\providecommand{\BIBentryALTinterwordspacing}{\spaceskip=\fontdimen2\font plus
\BIBentryALTinterwordstretchfactor\fontdimen3\font minus
  \fontdimen4\font\relax}
\providecommand{\BIBforeignlanguage}[2]{{%
\expandafter\ifx\csname l@#1\endcsname\relax
\typeout{** WARNING: IEEEtran.bst: No hyphenation pattern has been}%
\typeout{** loaded for the language `#1'. Using the pattern for}%
\typeout{** the default language instead.}%
\else
\language=\csname l@#1\endcsname
\fi
#2}}
\providecommand{\BIBdecl}{\relax}
\BIBdecl

\bibitem{URL_toyota}
\BIBentryALTinterwordspacing
P.~Koopman, ``{A Case Study of Toyota Unintended Acceleration and Software
  Safety},'' Oct. 2014, \textit{Accessed 2017-01-14}. [Online]. Available:
  \url{http://betterembsw.blogspot.se/2014/09/a-case-study-of-toyota-unintended.html}
\BIBentrySTDinterwordspacing

\bibitem{URL_tesla}
\BIBentryALTinterwordspacing
T.~Motors, ``{Tesla Autopilot},'' 2016, \textit{Accessed 2017-01-14}. [Online].
  Available: \url{http://www.tesla.com/presskit/autopilot#autopilot}
\BIBentrySTDinterwordspacing

\bibitem{VW16}
\BIBentryALTinterwordspacing
S.~V\"{o}st and S.~Wagner, ``Towards continuous integration and continuous
  delivery in the automotive industry,'' 2016. [Online]. Available:
  \url{arxiv.org/abs/1612.04139v1}
\BIBentrySTDinterwordspacing

\bibitem{FGMM16}
F.~Fagerholm, A.~S. Guinea, H.~M{\"a}enp{\"a}{\"a}, and J.~M{\"u}nch, ``The
  right model for continuous experimentation,'' \emph{Journal of Systems and
  Software}, 2016.

\bibitem{GYBC16}
F.~Giaimo, H.~Yin, C.~Berger, and I.~Crnkovic, ``{Continuous Experimentation on
  Cyber-Physical Systems: Challenges and Opportunities},'' in \emph{XP '16
  Workshops: Proceedings of the Scientific Workshop Proceedings of
  XP2016}.\hskip 1em plus 0.5em minus 0.4em\relax Edinburgh, Scotland UK: ACM,
  May 2016.

\bibitem{URL_volvo}
\BIBentryALTinterwordspacing
V.~Cars, ``{Drive Me Pilot},'' 2016, \textit{Accessed 2017-01-14}. [Online].
  Available:
  \url{http://www.volvocars.com/intl/about/our-innovation-brands/intellisafe/autonomous-driving/drive-me}
\BIBentrySTDinterwordspacing

\bibitem{URL_revere}
\BIBentryALTinterwordspacing
``{ReVeRe - Research Vehicle Resource at Chalmers},'' \textit{Accessed
  2017-01-14}. [Online]. Available:
  \url{https://www.chalmers.se/safer/EN/projects/pre-crash-safety/projects/revere-research-vehicle}
\BIBentrySTDinterwordspacing

\bibitem{URL_gcdc2011}
\BIBentryALTinterwordspacing
{High Tech Automotive Systems}, ``{Grand Cooperative Driving Challenge},'' Jan.
  2011, \textit{Accessed 2017-01-14}. [Online]. Available:
  \url{www.gcdc.net/en}
\BIBentrySTDinterwordspacing

\bibitem{BD08}
C.~R. Baker and J.~M. Dolan, ``Traffic interaction in the urban challenge:
  Putting boss on its best behavior,'' in \emph{2008 IEEE/RSJ International
  Conference on Intelligent Robots and Systems}.\hskip 1em plus 0.5em minus
  0.4em\relax IEEE, 2008, pp. 1752--1758.

\bibitem{BT16}
\BIBentryALTinterwordspacing
S.~Behere and M.~T\"{o}rngren, ``A functional architecture for autonomous
  driving,'' in \emph{Proceedings of the First International Workshop on
  Automotive Software Architecture}, ser. WASA '15.\hskip 1em plus 0.5em minus
  0.4em\relax New York, NY, USA: ACM, 2015, pp. 3--10. [Online]. Available:
  \url{http://doi.acm.org/10.1145/2752489.2752491}
\BIBentrySTDinterwordspacing

\bibitem{BD14}
C.~Berger and M.~Dukaczewski, ``Comparison of architectural design decisions
  for resource-constrained self-driving cars-a multiple case-study.'' in
  \emph{GI-Jahrestagung}, 2014, pp. 2157--2168.

\bibitem{URL_avoidinternet}
\BIBentryALTinterwordspacing
J.~Condliffe, ``{Why Some Autonomous Cars Are Going to Avoid the Internet},''
  \textit{Accessed 2017-01-14}. [Online]. Available:
  \url{http://www.technologyreview.com/s/603339/why-some-autonomous-cars-are-going-to-avoid-the-internet}
\BIBentrySTDinterwordspacing

\bibitem{Broy06}
\BIBentryALTinterwordspacing
M.~Broy, ``Challenges in automotive software engineering,'' in
  \emph{Proceedings of the 28th International Conference on Software
  Engineering}, ser. ICSE '06.\hskip 1em plus 0.5em minus 0.4em\relax New York,
  NY, USA: ACM, 2006, pp. 33--42. [Online]. Available:
  \url{http://doi.acm.org/10.1145/1134285.1134292}
\BIBentrySTDinterwordspacing

\bibitem{URL_docker}
``{Docker},'' https://www.docker.com, \textit{Accessed 2017-01-14}.

\bibitem{URL_opendavinci}
\BIBentryALTinterwordspacing
``{Open Source Development Architecture for Virtual, Networked, and
  Cyber-Physical System Infrastructures},'' \textit{Accessed 2017-01-14}.
  [Online]. Available: \url{http://www.opendavinci.org}
\BIBentrySTDinterwordspacing

\bibitem{MTA+16}
P.~Masek, M.~Thulin, H.~Andrade, C.~Berger, and O.~Benderius, ``Systematic
  evaluation of sandboxed software deployment for real-time software on the
  example of a self-driving heavy vehicle,'' in \emph{Proceedings of the 19th
  International Conference on Intelligent Transportation Systems (ITSC)}.\hskip
  1em plus 0.5em minus 0.4em\relax IEEE, 2016.

\bibitem{URL_protobuf}
``{Google Protocol Buffers},'' https://github.com/google/protobuf,
  \textit{Accessed 2017-01-14}.

\bibitem{URL_misra}
\BIBentryALTinterwordspacing
``{Motor Industry Software Reliability Association},'' \textit{Accessed
  2017-01-14}. [Online]. Available: \url{https://www.misra.org.uk/}
\BIBentrySTDinterwordspacing

\bibitem{URL_iso26262}
\BIBentryALTinterwordspacing
``{ISO 26262-1:2011 ``Road vehicles - Functional safety''},'' \textit{Accessed
  2017-01-14}. [Online]. Available:
  \url{http://www.iso.org/iso/catalogue_detail?csnumber=43464}
\BIBentrySTDinterwordspacing

\bibitem{URL_astazero}
\BIBentryALTinterwordspacing
``{AstaZero -- The world’s first full-scale test environment for future road
  safety},'' \textit{Accessed 2017-01-14}. [Online]. Available:
  \url{http://www.astazero.com}
\BIBentrySTDinterwordspacing

\bibitem{URL_copplar}
\BIBentryALTinterwordspacing
``{COPPLAR Project - CampusShuttle cooperative perception and planning
  platform},'' \textit{Accessed 2017-01-14}. [Online]. Available:
  \url{https://www.chalmers.se/safer/EN/projects/pre-crash-safety/projects/copplar-campusshuttle}
\BIBentrySTDinterwordspacing

\end{thebibliography}

\end{document}